\documentstyle[prl,aps,psfig,epsf]{revtex}
\begin{document}
\draft

\title{Decay of neutron-rich Mn nuclides and deformation of heavy Fe isotopes}

\author{
M. Hannawald$^1$, T. Kautzsch$^1$, A. W\"ohr$^2$, W.B. Walters$^3$, K.-L. Kratz$^1$,
V.N. Fedoseyev$^4$, V.I. Mishin$^4$,   
W. B\"ohmer$^1$, B. Pfeiffer$^1$, V. Sebastian$^5$, Y. Jading$^6$,  U. K\"oster$^7$,
J. Lettry$^6$, H.L. Ravn$^6$, and the ISOLDE Collaboration$^6$}

\address{
$^1$Institut f\"ur Kernchemie, Universit\"at Mainz,  D-55099 Mainz, Germany
}
\address{
$^2$Instituut voor Kern- en Strahlingsfysica, University of Leuven, B-3001
Leuven, Belgium}
\address{
$^3$Department of Chemistry, University of Maryland, College Park, MD 20742, 
USA}
\address{
$^4$Institute of Spectroscopy, Russian Academy of Sciences, RUS-142092 Troitzk, 
Russia}
\address{
$^5$Institut f\"ur Physik, Universit\"at Mainz, D-55099 Mainz, Germany} 
\address{$^6$CERN, CH-1211 Geneva 23, Switzerland}
\address{
$^7$Physik-Department, TU M\"unchen, D-85748 Garching, Germany}

\date{\today}
\maketitle

\begin{abstract}
The use of chemically selective laser ionization combined with 
$\beta$-delayed neutron counting at CERN/ISOLDE has permitted identification and 
half-life measurements for 623-ms $^{61}$Mn up through 14-ms $^{69}$Mn. 
The measured half-lives are found to be significantly longer near N=40 
than the values calculated with a QRPA shell-model using ground-state 
deformations from the FRDM and ETFSI models.
Gamma-ray singles and coincidence spectroscopy has been performed for 
 $^{64,66}$Mn decays to levels of $^{64,66}$Fe, revealing a significant
drop in the energy of the first 2$^+$  state in these nuclides that 
suggests an unanticipated increase in collectivity near N=40. 

\end{abstract}


\narrowtext

\twocolumn

Until recently, the principal data available for neutron-rich nuclides between 
$^{48}$Ca and $^{78}$Ni were the $\gamma$-spectroscopic data obtained in the
 1980's at GSI \cite{1,2}  and the nuclear masses reported by Seifert 
et~al. \cite{3}. During the past three years, however, a number of new 
experimental studies on level structures and decay 
properties on Fe-group nuclei have been performed \cite{4,5,6,7,8,9}. On the 
theoretical side, nuclear masses, ground-state (g.s.) deformations
and $\beta$-decay properties were calculated by M\"oller et al. 
\cite{10} on the basis of the FRDM mass model and the 
QRPA shell-model. Aboussir et al. \cite{11} also calculated 
masses and g.s.-deformations on the basis of the ETFSI-1 approach that 
are not always in agreement with those of the FRDM for the Fe-group 
nuclides considered here. 
Earlier, Richter et al. \cite{12} had performed fully microscopic shell-model 
calculations in this mass region and significantly underpredicted the measured
\cite{3} binding 
energies for the Cr to Fe isotopes with N$>$36. In addition to the clear 
nuclear-structure interest,  the neutron-rich Fe-group nuclei may also play 
an important role as possible seed nuclei in the astrophysical r-process 
\cite{13}. In the present paper, we report new measurements for the 
half-lives of heavy Mn nuclides up to $^{69}$Mn and for the level structure 
of $^{64,66}$Fe populated in the decays of $^{64,66}$Mn.

Manganese isotopes were produced at CERN by 1-GeV proton-induced 
spallation of uranium in a thick  UC$_2$ target at the ISOLDE facility. 
The ionisation of the Mn atoms was accomplished using a chemically 
selective, three-step laser resonance excitation scheme as described  
in detail earlier \cite{14}.

 Beams of Mn nuclides with masses differing by 
$\Delta$A$\ge$4 were transported separately to two different beam lines 
equipped with moving tape systems where ${\beta}$-delayed neutron (d.n.) 
multiscaling and $\gamma$-ray singles and coincidence measurements could 
be performed independently. In both cases, counting took place directly 
at the point of deposit, and the tape systems were used to remove the 
daughter nuclides as well as unavoidable surface-ionized isobaric Ga 
activities. Because the Mn half-lives being sought are in the millisecond 
range, data acquisition in both systems was initiated by the CERN-PSB 
proton pulses, separated by a multiple of 1.2~s, and continued for 1.0~s 
for each cycle.

Beta-delayed neutron data of high statistical quality were collected by 
multiscaling measurements using the Mainz $4\pi$$^3$He neutron counter. The 
time dependence of the counting rates for  $^{65-69}$Mn is 
shown in Figure~1. The decay curves were fitted with a constant small 
d.n.-background component up through A=65. Because there exist no measured 
d.n.-emission probabilities (P$_n$-values)  for the A$>$65 daughter and grand-daughter 
isobars, for the fits of 
the heavier 
isotopes were performed using theoretical P$_n$-values \cite{10} along with 
the known half-lives \cite{5,7,8,15}. For A=66 to 68, the contributions from d.n.-emission of the Fe 
and Co isobars are quite small and actually do not affect the Mn half-life 
fits. For A=69, however, a multi-component fit was necessary to account for 
the significant Fe and Co d.n.-branches. The resulting data are 
summarized in Table 1, and are compared to literature values and QRPA 
predictions using experimental masses 
as far as they are available \cite{15} and g.s.-deformations of the 
${\beta}$-decay daughter isotopes from the FRDM \cite{10} 
and ETFSI \cite{11} models.

For $^{61-63}$Mn, the half-lives that we have observed are somewhat 
shorter than the literature values and have considerably smaller 
uncertainties. In a recent report of a parallel experiment at Ganil, 
Sorlin et al. \cite{5} found half-lives similar to ours for $^{64-66}$Mn. However, neither our nor their data are in 
agreement 
 with those  reported by Ameil et al. at ENAM'95 
\cite{7} and cited in NUBASE \cite{15}. In a subsequent publication the 
same authors report somewhat shorter half-lives for these nuclides 
\cite{16}, however still exceeding 
our and Sorlin's values. Similarly, Franchoo et al. found systematic 
differences between their Ni half-lives \cite{8} and those reported 
in \cite{7,16}. These deviations gain particular importance when 
considering the conclusion of Ameil et al. that recent theoretical 
half-lives are not an improvement over calculations made almost a 
decade ago \cite{16}.

A comparison of the experimental Mn half-lives  with the 
predictions derived from QRPA calculations of Gamow-Teller (GT) strength 
functions \cite{17} (see Table~1) indicates that the theoretical half-lives for the 
g.s.-decays of $^{61-66}$Mn are on the average shorter than the measured 
ones by a factor of 2.6, whereas for the heaviest isotopes $^{67-69}$Mn 
the agreement becomes gradually better.
 When looking in more detail into 
the theoretical $\beta$-strength distributions, it becomes evident that in 
the decay of all neutron-rich Mn isotopes considered here, the (low-lying)
$\nu$f$_{5/2}$$\rightarrow$$\pi$f$_{7/2}$ transition is strongly 
dominating the GT-decay with I$_{\beta}$$\simeq$85--95$\%$ and  
log(ft)$\simeq$4.0, practically independent of the assumed g.s.-shape 
of the Fe daughters \cite{10,11}. As will be discussed later, this 
GT-pattern is, indeed, observed in the $\gamma$-data of $^{64}$Mn and 
$^{66}$Mn decay. With this rather ''simple'' decay pattern, already the 
differences between the experimental and theoretical half-lives seem to 
reflect the actual strength of this specific spin-flip transition.   

The $\gamma$-ray data were written in event-by-event mode for $\beta$-gated 
$\gamma$-singles as well as $\gamma\gamma$(t) coincidences with time 
recorded relative to each proton pulse. In this way, consecutive spectra 
of variable time intervals could be reconstructed from the data. As 
considerable data exist for the 
structure of even-even Fe nuclides up to A=62, in the present study we 
have focused on the $\gamma$-spectra for decay of 89-ms  $^{64}$Mn and 
66-ms $^{66}$Mn.  

At A=64, $\gamma$-ray peaks up to 4.2 MeV could be assigned to the decay of 
$^{64}$Mn. More than 20 lines have been incorporated into a decay scheme of 
at least 8 excited levels \cite{18}. In the upper part of Figure 2 are 
shown partial $\gamma$-ray spectra, one for the time period from 40~ms 
to 140~ms after the PSB proton pulses, and a second for the time slice 
from 800~ms to 900~ms after bombardment. The most intense line in the early 
$^{64}$Mn spectrum that decays with a short half-life is at 746~keV. With 
the intensity of this $\gamma$-line being more than five-times stronger 
than the next most intense peak, it is taken to be the 2$^+$ to 0$^+$ 
transition in the even-even daughter $^{64}$Fe. As five of the eight levels 
appear to depopulate to both the g.s. and the first 2$^+$ level, a low spin 
for the g.s. of $^{64}$Mn is indicated.  No candidate has so far been 
identified unambiguously for the 4$^+$ to 2$^+$ transition. On the basis 
of the observed $\gamma$$\gamma$-coincidences, strong GT-feeding to 
levels near 3.5~MeV in $^{64}$Fe is indicated \cite{18}, as predicted by our 
QRPA calculations. 

More than 20 $\gamma$-ray lines with energies up to at least 4.2 MeV have 
also been observed in the decay of 66-ms $^{66}$Mn and incorporated into
a partial level scheme containing 11 levels. We show in the 
lower part of Figure 2 a portion of the first and of a late spectrum 
(see above) between 525~keV and 900~keV. In the early spectrum, the 
strongest line by far is at 573~keV which is -- as in the above case of 
$^{64}$Fe -- taken as the g.s.-transition from the first 2$^+$ state in 
even-even $^{66}$Fe. 
The relatively weak $\gamma$ line at 840 keV (also shown in Figure 2) is found in coincidence with the 573-keV level and 
decays with the same half-life. Hence, it is assigned as depopulating the second excited state at 1414 keV and given a 
tentative spin and parity of 4$^+$.
A number of 
high-energy g.s.-transitions (E$>$2.7~MeV) again indicates strong 
GT-feeding to that energy region \cite{18}, as predicted by the QRPA. 

At both A=64 and 66, it was also possible to identify the growth and decay of
known daugther and granddaughter Fe and Co lines; 
whereas the lines from long-lived, surface-ionized  $^{64}$Ga and  
$^{66}$Ga showed no change in intensity  in our 1.0 s measuring period.

The 2$_1^+$ energies and the E$_4$/E$_2$ ratios for even-even 
$_{24}$Cr to $_{32}$Ge isotopes are 
shown in Figure 3.  As is well established, the 2$^+$ energy in the Z=28 
Ni nuclides rises sharply at N=40 and exhibits clear evidence for a 
semi-double shell closure, similar to that for Z=40, N=50 in $^{90}$Zr. 
Recent studies of the structure of $^{69}$Ni and $^{69}$Cu are consistent 
with the closed-shell character of $^{68}$Ni \cite{8,9}.

Raman et al. \cite{19} have presented an extensive discussion of the relationship between 2$^+$ energies, B(E2) values 
and collectivity. In particular, they have shown the inverse correlation between 2$^+$ energies and deformation. Similar 
correlations of B(E2) values with energy have recently been presented by Azaiez and Sorlin \cite{20}. Hence, the direct 
interpretation of 
the drop in 2$^+$ energy from 877 kev in $^{62}$Fe to 573 kev in $^{66}$Fe  would be an increase in deformation.
Because of the large quantity of data available, including the B(E2) systematics \cite{19}, there is general agreement 
on deformation values of $\beta_2\simeq0.18$ for the lighter Fe nuclides. For the heavier isotopes where only few data are 
available, both the FRDM \cite{10} and new calculations using the Relativistic Mean Field approach \cite{21} indicate 
deformation of $\sim$0.21 for $^{62}$Fe and then dropping toward values below 0.1 for $^{66}$Fe. In contrast, the ETFSI 
calculations \cite{11}, show a value of $\beta_2$=0.18 for $^{62}$Fe  that rises to 0.27 for $^{66}$Fe.

The data in Figure 3 
reveal that both the 2$_1^+$ energy and $4^+/2^+$ ratio for $^{66}$Fe$_{40}$ are comparable to those of 
stable $^{76}$Ge$_{44}$. 
Extensive studies of the structure of the stable Ge nuclides \cite{22,23} 
have indicated a range of deformation values between $\beta_2$=0.22 and 
0.28 for $^{76}$Ge.
Azaiez and Sorlin recently reported a $\beta_2$-value of 0.23 for $^{72}$Zn$_{42}$ that has a 2$^+$ energy of 653 keV 
\cite{20}.
Combining all of the above approaches, we 
deduce a $\beta_2\simeq0.26$ for $^{66}$Fe$_{40}$.
Thus, we conclude that the trends we observe  for the Fe isotopes are reproduced only by the ETFSI calculations \cite{11}.

We attribute this increase of deformation to the strong proton-neutron (pn) 
interaction 
between the two f$_{7/2}$ proton holes and  the g$_{9/2}$ neutrons, which 
results in a dramatic lowering of the energy of the $\nu$g$_{9/2}$ orbital.
The effect of interactions between protons and neutrons in high-j orbitals has also been discussed in \cite{8} 
for odd-mass Cu nuclides near N=40. In that paper, a sharp lowering of the energy of the $\pi$f$_{5/2}$ orbital was 
observed with increasing  
occupancy of the $\nu$g$_{9/2}$ state.
                                      
For the neutron-rich Fe and Mn nuclides, one effect of this strong pn-interaction is to lower the energy of the 
$\nu\mbox{g}_{9/2}$ orbital. For example, the recent report of a 9/2$^+$ M2 isomer in $^{61}$Fe$_{35}$ at 861 keV is well below the 
1292-keV position of the $\nu$g$_{9/2}$ state in isotonic $^{63}$Ni$_{35}$ \cite{9}. Because of the lowered position of the 
base  $\nu$g$_{9/2}$ orbital, it can be seen by reference to Figure~12 in \cite{10} or Figure~7 in [5b] that, as the 
neutron number increases beyond 36, the down-sloping $\nu[440]1/2^+$ and  $\nu[431]3/2^+$ orbitals are more likely to be 
occupied than the spherical orbitals thus generating increased deformation at N=40. This collectivity increase beyond 
N=36 can be seen to correlate with the difference between the masses calculated with a spherical model and the measured 
values shown by Richter et al. \cite{12} in their Figure~1. From the trends in that  figure, it can be expected that 
$^{64}$Cr$_{40}$ may be even more deformed than $^{66}$Fe$_{40}$.
                                      
Moreover, such a change in neutron occupancy near N=40 
may help to account for the  retarded Mn
$\beta$-decay rates described earlier. The decay of the odd-mass Mn 
nuclides would be considered as ``even-jumping'' transitions as described 
by Kisslinger and Sorensen \cite{24}. For such transitions, the 
$\beta$-decay rate $\lambda$ is proportional to 
(V$_n$$\cdot$V$_p$)$^2$$\cdot$M$^2$, where M$^2$ is the pure decay 
matrix element and V represents the proton/neutron occupancy values. As 
noted earlier, the allowed GT $\beta$-decay is dominated by 
the $\nu$f$_{5/2}\rightarrow\pi$f$_{7/2}$ transition.
Because of the increase in deformation, the negative parity orbitals $\nu[301]3/2^-$ and  $\nu[303]5/2^-$ with 
$\nu$f$_{5/2}$ parentage are up-sloping and are thus displaced from the region near the Fermi surface. Instead, the 37th through 40th neutrons 
will preferentially fill the down-sloping low-j, positive-parity orbitals with $\nu$g$_{9/2}$ origin that can only undergo 
forbidden $\beta$-decay.
The 
lowered occupancy of the above critical negative-parity states may well be 
directly responsible for the observed retardation of the $\beta$-decay 
rates for Mn nuclides with N$\simeq$40. Moreover, the ultimate increase in 
occupancy for these orbitals that must occur for  N$>$40  
also accounts for the gradual convergence of the calculated half-lives 
and the newly measured values  for 
$^{67-69}$Mn$_{42-44}$.
                                     
This work was supported by the German BMBF (06MZ864) and  DFG 
(436RUS/17/40/97 and Kr806/3), the Russian Foundation for Basic Research 
(96-02-18331), the Belgian FWO, and the U.S. DOE.

\begin{table}
\caption{  Experimental and theoretical half-lives for the neutron-rich Mn 
nuclides. In the QRPA calculations, deformation values $\beta_2$(Fe) 
were taken from FRDM [10] and ETFSI [11].}
\begin{tabular}{llllll} 
Mass &\multicolumn{5}{c}{Half-life [ms]}\\
     & This Work & Lit. & Ref.& \multicolumn{2}{c}{QRPA}\\
     &           &      &     &  [10] & [11] \\
61   & 623(10)   & 710(10)  & \cite{2}  & 234 & 231\\
62   & 671(5)    & 880(150) & \cite{15} & 267 & 274\\
63   & 275(4)    & 282(18)  & \cite{15} &  76 &  76\\
     &           & 321(22)  & \cite{5}  &     & \\
64   &  89(4)    & 240(30)  & \cite{7}  & 67  & 105\\
     &           & 140(30)  & \cite{16}  &     &\\
     &           &  91(7)   & \cite{5}  &     &\\
65   &  88(4)    & 160(30)  & \cite{7}  & 39  &46\\
     &           & 110(20)  & \cite{16}  &     &\\
     &           &  88(7)   & \cite{5}  &     &\\
66   &  66(4)    & 220(40)  & \cite{7}  & 21  &23\\
     &           &  90(20)  & \cite{16}  &     &\\
     &           &  62(13)  & \cite{5}  &     &\\
67   &  42(4)    &     &    &  25   &27\\
68   &  28(4)    &     &    &  18   &25\\
69   &  14(4)    &     &    &  16   &14\\
\end{tabular}
\end{table}

\begin{figure}
\centerline{\psfig{file=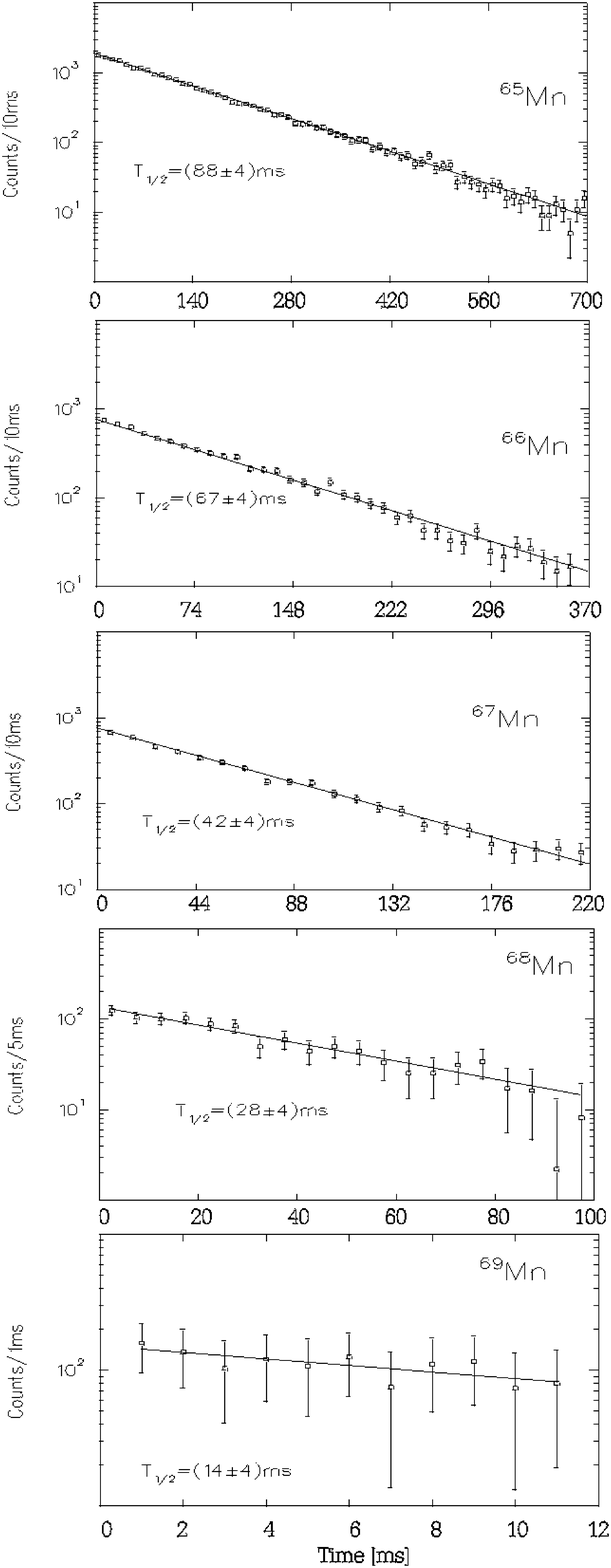,width=8.3cm}}
\caption{Beta-delayed neutron decay curves for the  neutron-rich isotopes $^{65-69}$Mn.}
\end{figure}

\begin{figure}
\centerline{\psfig{file=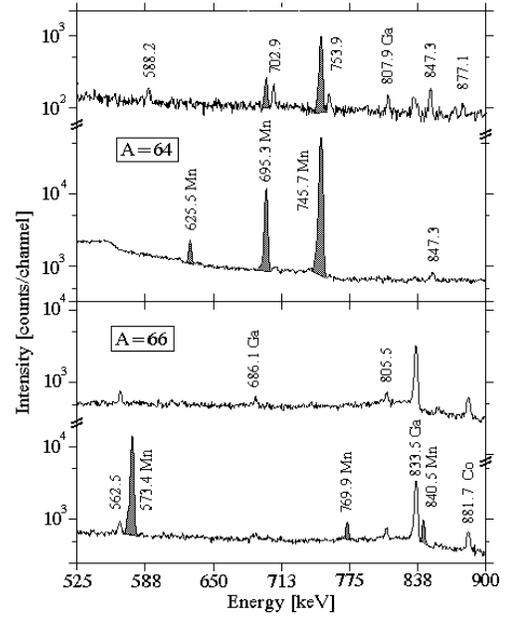,width=7.5cm}}
\caption{Partial $\gamma$-ray spectra taken at A=64 (upper part) and A=66 
(lower part), one for the time period from 40~ms to 140~ms after the PSB 
proton pulses, and a second for the time slice from 800~ms to 900~ms 
after bombardment.}
\end{figure}

\begin{figure}
\begin{center}
\begin{minipage}{7cm}
\psfig{file=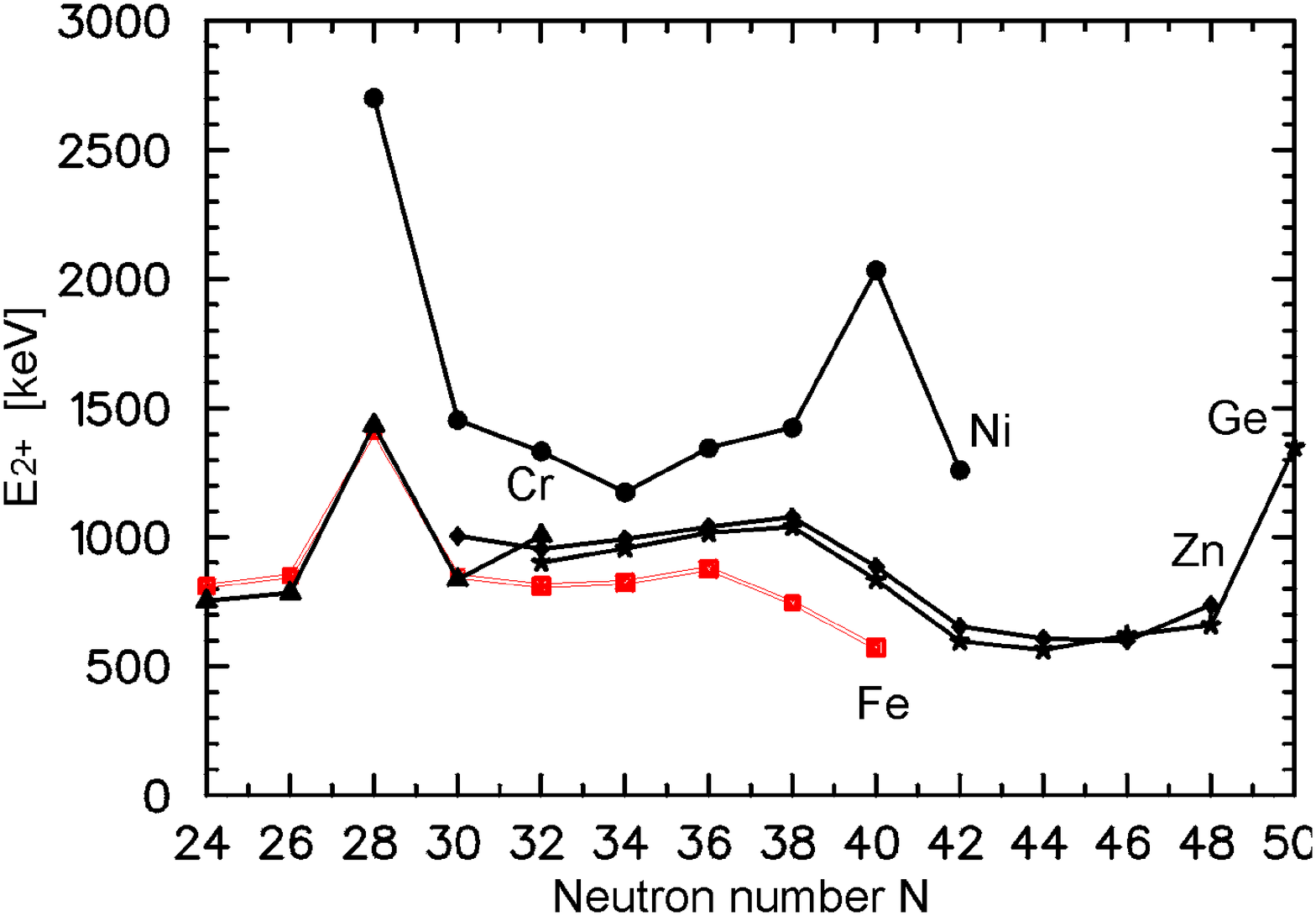,width=7cm,angle=0}
\end{minipage}
\begin{minipage}{7cm}
\psfig{file=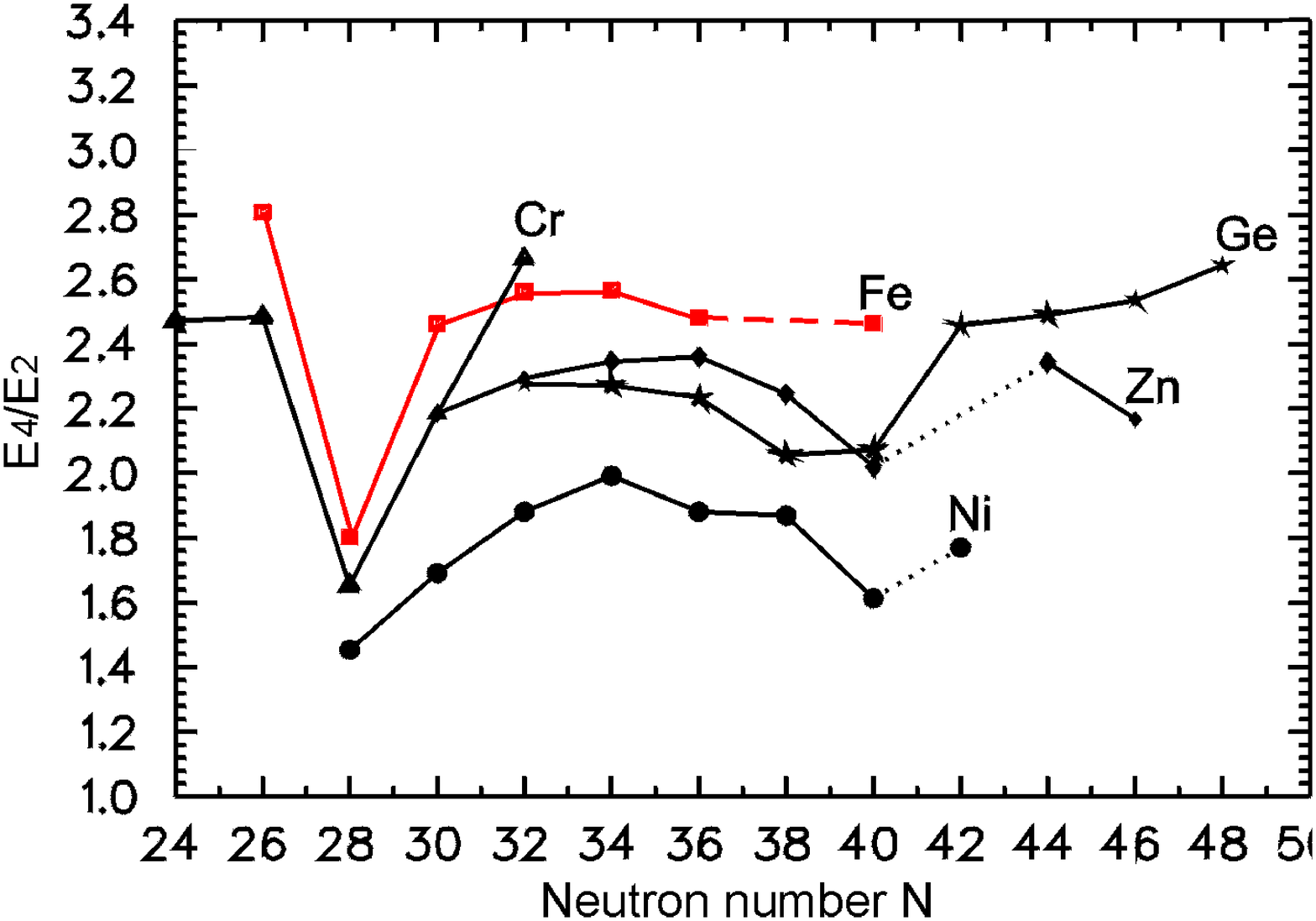,width=7cm,angle=0}
\end{minipage}
\end{center}
\caption{Energies of the first 2$^+$ levels and 
E$_{4^+}$/E$_{2^+}$ ratios of even-even $_{24}$Cr to $_{32}$Ge nuclides. }
\end{figure}

\end{document}